\documentclass[%
 reprint,
 showkeys,
 amsmath,amssymb,
 aps,
 prb,
 floatfix,
]{revtex4-2}

\usepackage[dvipsnames]{xcolor}
\usepackage{xurl}
\usepackage[colorlinks,allcolors=blue]{hyperref}
\hypersetup{colorlinks=true,allcolors=MidnightBlue}

\definecolor{Cu}{HTML}{B4B8BA}
\definecolor{Zn}{HTML}{4A70A1}
\definecolor{Sn}{HTML}{E0927D}
\definecolor{S}{HTML}{DFB352}
\definecolor{Ge}{HTML}{4CA8A1}
\definecolor{LMeditColor}{HTML}{4CA8A1}

\usepackage{colortbl}
\usepackage{booktabs}

\newcommand*{\tcw}{\color{white}}

\usepackage{braket}  
\usepackage{graphicx} 
\usepackage{dcolumn} 
\usepackage{multirow} 
\usepackage{bm}

\usepackage{xurl}
\usepackage[colorlinks,allcolors=blue]{hyperref}

\begin{document}

\preprint{APS/123-QED}

\title{Accurate bandgaps of 
photovoltaic kesterites
from first-principles DFT+U}

\author{Andrew C. Burgess$^{1,2,3}$}
\author{Lórien MacEnulty$^{1,4}$}
\author{Ethan D'Arcy$^{1}$}
\author{David Gavin$^{1,2}$}
\author{David D. O'Regan$^{1,2}$}
\email{david.o.regan@tcd.ie}
\affiliation{
 $^1$School of Physics, Trinity College Dublin, The University of Dublin, Ireland}
 \affiliation{
 $^2$CRANN Institute and AMBER Research Centre, Trinity College Dublin, The University of Dublin, Ireland}
 \affiliation{
 $^3$School of Physics, University College Dublin, Dublin 4, Ireland}
 \affiliation{
 $^4$Université catholique de Louvain, Institute of Condensed Matter and Nanoscience, Chemin des Étoiles 8, B-1348 Louvain-la-Neuve, Belgium}

\begin{abstract}
Streamlined prediction of  
the electronic properties 
of photoactive materials warrants a Density Functional Theory (DFT) based approach that {(i)} yields reliable bandgaps, {(ii)} is free of empirically tuned parameters, and {(iii)} exhibits low computational overhead. Here we show that for Cu$_2$ZnSnS$_4$ and Cu$_2$ZnGeS$_4$ kesterite photovoltaic materials, all three of these demands are met by the DFT plus Hubbard $U$ technique (DFT+$U$) with corrective parameters evaluated via minimum-tracking linear response. The predicted bandgaps are found to even marginally outperform those from the self-consistent GW approach. Key to this method's success is the application of Hubbard $U$ corrections to all atomic subspaces that dominate the conduction and valence band edges, as opposed to the conventional approach of correcting $3d$ and $4f$ atomic states. Intriguingly, the inclusion of Hund's $J$ corrections via the extended DFT+$U$+$J$ functional significantly worsens these results. This under performance can be ameliorated through the use of the Burgess-Linscott-O'Regan (BLOR) flat-plane based Hubbard $U$ plus Hund's $J$ functional, with bandgap predictions in close agreement with the  conventional DFT+$U$ method. The DFT+$U$ method is also used to predict  defect-induced changes to the bandgap and associated formation energies, in 1,728-atom supercells. 
\end{abstract}

\maketitle


The need to decarbonize the global economy is one of the greatest challenges faced by humanity in the twenty-first century. Both the European Union~\cite{Regulation20211119} and the United Kingdom~\cite{stewartUKsPlansProgress2025} are  committed to achieving net-zero greenhouse gas emissions by 2050. The wide-scale adoption of renewable energy technology will play a pivotal role in achieving this ambitious goal~\cite{EnergyRoadmap}. Photovoltaic (PV) materials have already been identified as key components of the wider renewable energy sector and are forecast to become the primary source of renewable electricity by 2030~\cite{renewables2024}. However, {owing} at least in part to the material's technological maturity, the commercial market is currently dominated by multicrystalline silicon-based PVs with a typical efficiency of $\sim20\%$ at most. Improving the efficiency and lowering the cost of PV technology thus demands the exploration of alternative materials.

\begin{figure}
\includegraphics[trim={0cm 0.9cm 0cm 0.6cm},clip,width=1.0\linewidth]{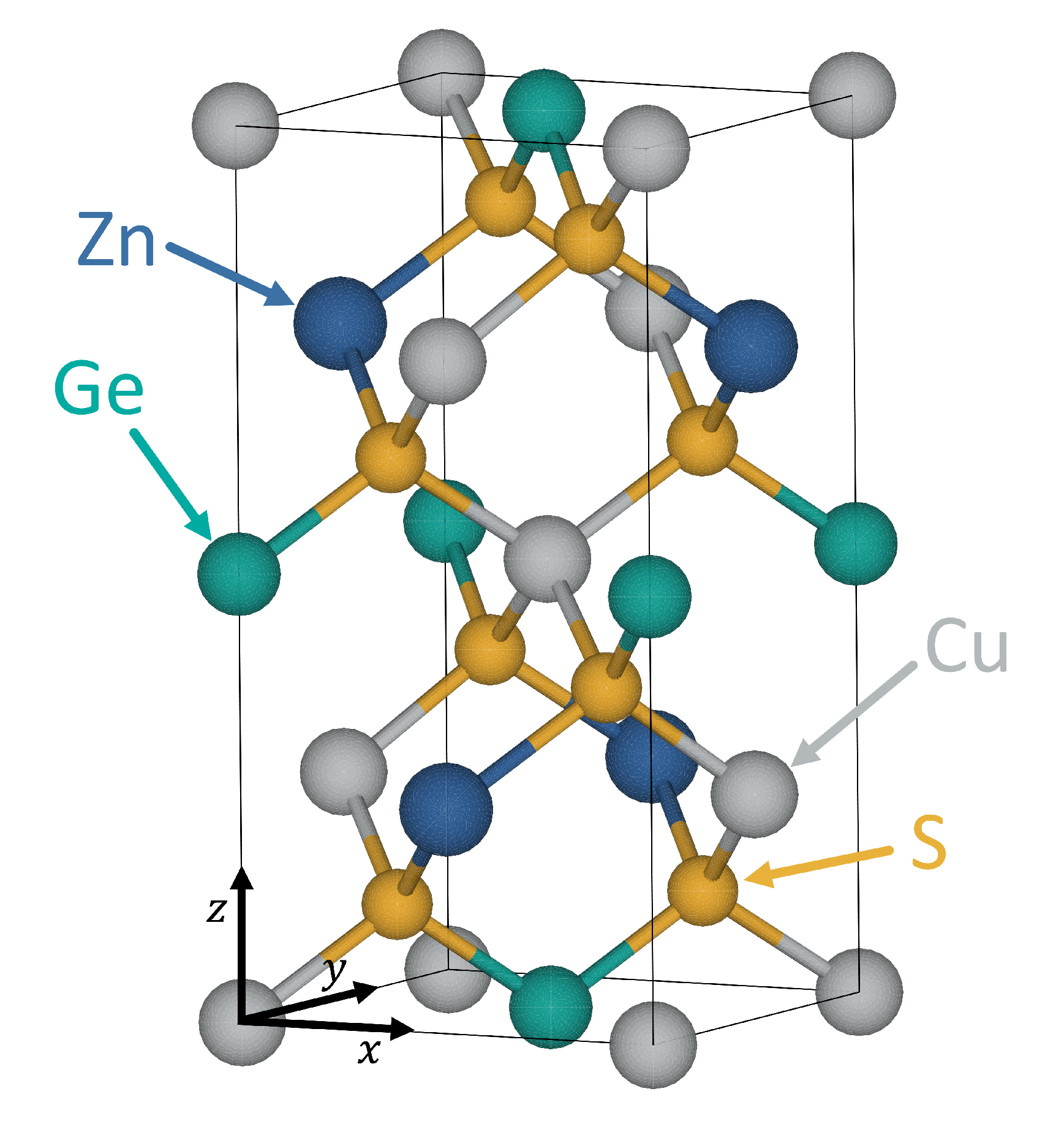} 
\caption{\label{KesteriteFigure} CZGS demonstrating the general crystal structure of a quaternary chalcogenide X$_2$Zn Y S$_4$ in the kesterite phase, space group $I\bar{4}$. The principle transition metal X=Cu is shown in gray, Zn is blue, Y=Sn,Ge in green, and S in yellow.}
\end{figure}

One class of promising candidates are the quaternary kesterite chalcogenides{—among them the prototypical Cu$_2$ZnSnS$_4$ (CZTS) and its relative, Cu$_2$ZnGeS$_4$ (CZGS; see Fig. \ref{KesteriteFigure})—}due to their {geographically} abundant and non-toxic elemental consituents~\cite{Hynes,USGS}, high absorption coefficients ($>10^4$ cm$^{-1}$)~\cite{katagiriPreparationEvaluationCu2ZnSnS41997} and near-optimal single junction bandgap of $\sim1.5$ eV~\cite{seolElectricalOpticalProperties2003}. Furthermore, similarities between fabrication techniques of these materials and CIGS (CuIn$_x$Ga$_{1-x}$Se$_2$)~\cite{sureshPresentStatusSolutionProcessing2021}, their more advanced yet more expensive cousins, ensures that engineers can leverage existing industrial infrastructure to expedite their entry into the PV market~\cite{Schnabel2017,Kohl2018}.

Despite periods of rapid improvement~\cite{katagiriDevelopmentCZTSbasedThin2009,wangDeviceCharacteristicsCZTSSe2014} since the seminal work of Ito and Nakazawa~\cite{nakazawaElectricalOpticalProperties1988}, the record efficiency of kesterite-based solar cells remains low, at 15.8\%~\cite{zhouControlPhaseEvolution2023,greenSolarCellEfficiency2025}. {This efficiency reaches less than half of the predicted Shockley-Queisser limit for these materials \cite{Ki2011}. Fortunately, kesterite bandgap energies have been shown to be quite cleanly manipulated by partial doping or alloy substitution~\cite{Polman2016,Dhruba2015,Qi2017,Giraldo2019,heinrichEffectIsovalentSubstitution2014}.} Unlocking improved cell efficiencies thus requires a deeper understanding of the electronic and atomistic structure of quaternary kesterite chalcogenides, {an understanding that is potentially accessible through density functional theory (DFT) calculations}. However, conventional {approximate DFT calculations} with local and semi-local exchange correlation functionals erroneously predict CZTS as being near metallic in character, with a bandgap of less than 0.25 eV~\cite{wexlerExchangecorrelationFunctionalChallenges2020,paier$textCu_2textZnSnS_4$PotentialPhotovoltaic2009}. Improved bandgap predictions, as {best-practices among modelers of photoactive materials demand}, can be achieved through deployment of hybrid functionals such as that of Heyd-Scuseria-Ernzerhof, (HSE)~\cite{heydHybridFunctionalsBased2003}, albeit usually with a considerable increase in computational cost. Such an increase in cost makes the approach unsuitable for use in large simulation cells, {which are required for accurate modeling of defects,} or when sampling many configurations. 
The Hubbard $U$-corrected DFT (DFT+$U$) approach~\cite{himmetogluHubbardcorrectedDFTEnergy2014,anisimovFirstprinciplesCalculationsElectronic1997,kirchner-hallExtensiveBenchmarkingDFT2021} offers a pragmatic alternative capable of yielding reliable bandgap predictions at comparable cost to the (semi-)local base functional that it is designed to correct.

Three DFT$+U$-type corrective functionals have been assessed in this study, namely the well-established DFT$+U$ functional of Dudarev \textit{et al.}~\cite{dudarevElectronenergylossSpectraStructural1998}, the extended DFT+$U$+$J$ functional of Himmetoglu \textit{et al.}~\cite{himmetogluFirstprinciplesStudyElectronic2011} and the recently developed Burgess-Linscott-O'Regan (BLOR) functional~\cite{burgessMathrmDFTTexttypeFunctional2023}. The corrective functional of Dudarev \textit{et al.} was designed to explicitly treat the effective electron-electron interaction within a selected subspace using an on-site interaction term inspired by the Hubbard model,
\begin{equation}
E_{\rm int}=\frac{U}{2}\sum_{\sigma,  m, m'}n^{\sigma}_{mm} n^{\bar{\sigma}}_{m'm'}+\frac{U-J}{2}\sum_{\sigma,  m \neq m'}n^{\sigma}_{mm} n^{\sigma}_{m'm'},
\end{equation}
where $U$ and $J$ are the subspace-averaged, screened Coulomb and exchange interactions—{typically called the Hubbard $U$ and Hund's $J$ parameters and to which we will collectively refer as the Hubbard parameters—}and $n^{\sigma}_{mm}$ is the spin-$\sigma$ occupancy of orbital $m$ in the selected subspace. However, in this model, said electron-electron interaction is assumed to have already been accounted for to a less favorable extent by the approximate Hartree and exchange correlation (Hxc) functional, which the DFT+$U$ functional is designed to supplement. Dudarev \textit{et al.} mitigate this issue by invoking a spin-polarized, fully-localized limit-type double counting correction to arrive at the following expression for their DFT$+U$-functional \footnote{In the derivation of their DFT$+U$ functional, Dudarev \textit{et al.} include an additional minor adjustment to ensure their functional form is invariant under unitary transformations of the subspace orbitals.},
\begin{equation}
E_{\rm u}=\frac{U-J}{2}\sum_{\sigma,  m, m'}n^{\sigma}_{mm'}\delta_{m'm'}-n^{\sigma}_{mm'}n^{\sigma}_{m'm},
\label{eqn:dudarev}
\end{equation}
where the occupancy matrix elements can be defined using the set of subspace orbitals $\{\phi_m \}$ and the spin-resolved Kohn-Sham {(KS)} density operator $\hat{\rho}^{\sigma}$,
\begin{equation}
n^{\sigma}_{mm'}=\braket{\phi_m|\hat{\rho}^{\sigma}|\phi_{m'}}.
\end{equation}
The $U-J$ prefactor in Eq.~\ref{eqn:dudarev} is typically labelled as the effective Hubbard parameter $U_{\rm eff}$. Like the DFT+$U$ functional of Dudarev \textit{et al.}, the DFT+$U$+$J$ functional of Himmetoglu \textit{et al.} is based on an on-site interaction term and a spin-polarized, fully-localized-limit double-counting correction. However, in this case, the interaction term is derived from Hartree-Fock theory, keeping only the on-site Coulomb and exchange terms, which are treated at the subspace averaged level, yielding 
\begin{widetext}
\begin{equation}
 E_{\rm int}=\frac{U}{2}\sum_{\sigma}\left(n^{\sigma}n^{\sigma}
+n^{\sigma}n^{\bar{\sigma}}\right)-\frac{U}{2}\sum_{\sigma,m,m'}n^{\sigma}_{mm'}n^{\sigma}_{m'm} 
+\frac{J}{2}\sum_{\sigma,m,m'}\left(n^{\sigma}_{mm'}n^{\sigma}_{m'm}+n^{\sigma}_{mm'}n^{\bar{\sigma}}_{m'm}\right)-\frac{J}{2}\sum_{\sigma}n^{\sigma}n^{\sigma}. 
\end{equation}
\end{widetext}
Subtracting off the double-counting correction and omitting the numerically fraught minority-spin term results in the 
simplified rotationally-invariant DFT+$U$+$J$ functional given by 
\begin{align}
E_{\rm u}=&\frac{U-J}{2}\sum_{\sigma,  m, m'}n^{\sigma}_{mm'}\delta_{mm'}-n^{\sigma}_{mm'}n^{\sigma}_{m'm}\nonumber \\ &+\frac{J}{2}\sum_{\sigma,  m, m'}n^{\sigma}_{mm'}n^{\bar{\sigma}}_{m'm}.
\end{align}

In contrast to invoking an on-site interaction term and a double counting correction, 
the third functional 
considered there, the BLOR functional, is specifically designed to enforce the flat plane condition on localized subspaces, an exact condition of DFT which defines the shape of the total energy surface as a function of the electron count and spin-magnetization~\cite{yangDegenerateGroundStates2000,mori-sanchezDiscontinuousNatureExchangeCorrelation2009}. By virtue of enforcing the flat plane condition, the BLOR functional by design corrects for the subspace analogue of many-electron self-interaction error~\cite{ruzsinszkySpuriousFractionalCharge2006,mori-sanchezManyelectronSelfinteractionError2006} and static correlation error~\cite{cohenFractionalSpinsStatic2008}, two of the most pervasive errors of DFT, which can be defined as energetic deviations from the flat plane condition with respect to electron count and spin-magnetization, respectively. A many-body generalisation of the BLOR functional~\cite{burgessFlatPlaneBased2024} and the DFT+$U$ functional of Bajaj et al.~\cite{bajajCommunicationRecoveringFlatplane2017,bajajNonempiricalLowcostRecovery2019} have also been developed based on the flat-plane condition but are beyond the scope of this study.

The BLOR corrective functional has two different forms depending on whether the subspace is more or less than half-occupied, i.e., the total subspace occupancy $N>2l+1$ or $N\leq 2l+1$, where $l$ is the orbital angular momentum quantum number. In terms of subspace occupancy matrix elements the corrective functional is
\begin{widetext}
\begin{align}
\label{eqn:BLOR_sum_version}
E_{\rm BLOR}= \left\{
\begin{array}{*8{>{\displaystyle}c}}
\sum_{\sigma m m'}\frac{U^{\sigma}}{2}n^{\sigma}_{mm'}\delta_{mm'}
&-&
\frac{U^{\sigma}}{2}n^{\sigma}_{mm'}n^{\sigma}_{m'm}
&-&
\frac{U^{\sigma}+2J}{2}n^{\sigma}_{mm'}n^{\bar{\sigma}}_{m'm},
&\bf{}&
\bf{}
&  \
N \leq 2l+1,
 \\
\sum_{\sigma m m'}  \bigg(U^{\sigma}+\frac{U^{\bar{\sigma}}}{2}+2J\bigg)n^{\sigma}_{mm'}\delta_{mm'}
&-&
\frac{U^{\sigma}}{2}n^{\sigma}_{mm'}n^{\sigma}_{m'm}
&-&
\frac{U^{\sigma}+2J}{2}n^{\sigma}_{mm'}n^{\bar{\sigma}}_{m'm}
&-&
\frac{U^{\sigma}+2J}{2(2l+1)},
&  \
N > 2l+1.
\end{array}
\right.
\end{align}
\end{widetext}
The BLOR functional includes a spin-dependent Hubbard parameter $U^{\sigma}$,  however for non-spin polarized systems such as CZTS, the spin-indexed parameter $U^{\sigma}=U-J$. This article is the first reported application of the BLOR functional to solid-state systems.

Transition metal $3d$ and Lanthanide $4f$ atomic orbitals are the most commonly selected subspaces for treatment at the DFT$+U$ level. However, the approach has been extended to a wide variety of other atomic orbitals such as oxygen $2p$~\cite{mooreHighthroughputDeterminationHubbard2024a,consiglio2016importance,yang2025dft+}, transition metal $4s$~\cite{lambertUseMathrmDFT2023,kulik2010systematic} and nearest neighbor ligand states~\cite{macenultyBenchmarkingTotalEnergies2025}. Alternative subspace definitions such as maximally-localized Wannier functions~\cite{cartaExplicitDemonstrationEquivalence2025,shihScreenedCoulombInteractions2012,fabrisTamingMultipleValency2005}, ortho-atomic orbitals~\cite{lambertUseMathrmDFT2023,timrovPulayForcesDensityfunctional2020,riccaSelfconsistentDFTStudy2020,timrovElectronicStructurePristine2020}, and molecular orbitals~\cite{bajajMolecularDFT+UTransferable2021,bajajMolecularOrbitalProjectors2022} have also been thoroughly investigated. 

Once a specific DFT+$U$ functional and targeted subspace has been selected, the $U$ and $J$ parameters must be selected. Unfortunately, it remains commonplace in the literature to empirically tune the Hubbard corrective parameters to reproduce material properties of interest, severely limiting the predictive power of the DFT$+U$ approach. By contrast, in this study, we deploy the minimum-tracking linear response methodology~\cite{linscottRoleSpinCalculation2018,moynihanSelfconsistentGroundstateFormulation2017,orhanFirstprinciplesHubbardHund2020,bermanReconcilingTheoreticalExperimental2023} to evaluate Hubbard $U$ and Hund's $J$ corrective parameters for select atomic subspaces. This is achieved by applying a series of spin-resolved perturbations to the subspace of interest and recording the change in spin-resolved subspace occupancy $n^{\sigma}$ and spin-resolved, subspace-averaged Hxc potential $v_{\rm Hxc}^{\sigma}$ where,
\begin{equation}
 n^{\sigma} =\sum_m  n^{\sigma}_{mm} \quad \& \quad v_{\rm Hxc}^{\sigma}=\frac{\sum_m\braket{\phi_m|\hat{v}^{\sigma}_{\rm Hxc}|\phi_{m}}}{\sum_{m'}\braket{\phi_{m'}|\phi_{m'}}}.
\end{equation}
Within this formalism, the Hubbard $U$ and Hund's $J$ parameters are defined as
\begin{equation}
\label{eqn:basicUandJ}
U=\frac{1}{2}\frac{dv_{\rm Hxc}^{\upharpoonright}+dv_{\rm Hxc}^{\downharpoonright}}{d(n^{\upharpoonright}+n^{\downharpoonright})} \quad \& \quad J=-\frac{1}{2}\frac{dv_{\rm Hxc}^{\upharpoonright}-dv_{\rm Hxc}^{\downharpoonright}}{d(n^{\upharpoonright}-n^{\downharpoonright})}.
\end{equation}
The definition of the Hubbard $U$ parameter can be re-expressed in terms of the elements of the spin-indexed, subspace-averaged Hxc kernel,
\begin{equation}
\label{eqn:hubbard_u_in_terms_of_fhxc}
U\approx\frac{1}{2}\frac{f^{\upharpoonright \upharpoonright}\delta n^{\upharpoonright} +f^{\upharpoonright \downharpoonright}\delta n^{\downharpoonright}+f^{\downharpoonright \upharpoonright}\delta n^{\upharpoonright}+f^{\downharpoonright \downharpoonright}n^{\downharpoonright}}{\delta(n^{\upharpoonright}+n^{\downharpoonright})}.
\end{equation}
To proceed further, one of two possible approximations needs to be made, i.e., the simple $2 \times 2$ approach or the scaled $2 \times 2$ approach. However, these are exactly equivalent  for nonmagnetic materials, which are the exclusive focus of this study. Within the simple $2 \times 2$ approach, it is assumed that the spin-resolved occupancies respond equally to a non-spin-polarized perturbation, i.e., $\delta n^{\upharpoonright} \approx \delta n^{\downharpoonright} $ in Eq.~\ref{eqn:hubbard_u_in_terms_of_fhxc}, in which case the Hubbard {$U$ parameter simplifies to}
\begin{equation}
U=\frac{1}{4}\left(f^{\upharpoonright \upharpoonright}+f^{\upharpoonright \downharpoonright}+f^{\downharpoonright \upharpoonright}+f^{\downharpoonright \downharpoonright}\right).
\end{equation}
Analogously, the simple $2 \times 2$ approach for the Hund's parameter results in the expression
\begin{equation}
J=-\frac{1}{4}\left(f^{\upharpoonright \upharpoonright}-f^{\upharpoonright \downharpoonright}-f^{\downharpoonright \upharpoonright}+f^{\downharpoonright \downharpoonright}\right).
\end{equation}
The spin-indexed, subspace-averaged Hxc kernel
(a partially screened quantity, when calculated for subspaces of the 
global system) maybe expressed as
\begin{equation}
\label{eqn:Hxc_kernel}
\begin{pmatrix}
f^{\upharpoonright \upharpoonright} & f^{\upharpoonright \downharpoonright} \\
f^{\downharpoonright \upharpoonright} & f^{\downharpoonright \downharpoonright}
\end{pmatrix}=
\renewcommand\arraystretch{2}
\begin{pmatrix}
\frac{dv_{\rm Hxc}^{\upharpoonright}}{dv_{\rm  ext}^{\upharpoonright}} & \frac{dv_{\rm Hxc}^{\upharpoonright}}{dv_{\rm  ext}^{\downharpoonright}} \\ 
\frac{dv_{\rm Hxc}^{\downharpoonright}}{dv_{\rm  ext}^{\upharpoonright}}& \frac{dv_{\rm Hxc}^{\downharpoonright}}{dv_{\rm  ext}^{\downharpoonright}}
\end{pmatrix}
\renewcommand\arraystretch{2}
\begin{pmatrix}
\frac{dn^{\upharpoonright}}{dv_{\rm ext}^{\upharpoonright}} & \frac{dn^{\upharpoonright}}{dv_{\rm ext}^{\downharpoonright}}  \\ 
\frac{dn^{\downharpoonright}}{dv_{\rm ext}^{\upharpoonright}}  & \frac{dn^{\downharpoonright}}{dv_{\rm ext}^{\downharpoonright}} 
\end{pmatrix}^{-1},
\end{equation}
where $v_{\rm ext}^{\sigma}$ is the spin-$\sigma$, subspace-averaged external potential, typically defined relative to the ground state, in which case its value is equal to the spin-$\sigma$ perturbation strength. The matrix elements of $dv_{\rm Hxc}/dv_{\rm ext}$ and ${dn}/{dv_{\rm ext}}$ can be readily evaluated as the slopes of the corresponding linear response plots. The response of the Hxc contribution of the Projector Augmented Wave (PAW)~\cite{blochlProjectorAugmentedwaveMethod1994} effective potential need also be accounted for~\cite{macenultyBenchmarkingTotalEnergies2025} within Eq.~\ref{eqn:Hxc_kernel}, but in this study, norm-conserving pseudopotentials (NCPs) were deployed throughout.

For the avoidance of doubt, no inter-site response matrix
inversion was carried out, as this cumbersome step
is entirely unnecessary and
avoided in the minimum-tracking approach, when calculating
only on-site parameters, as we are.
In the present work, we also exploited the spin-symmetry 
of ultimately nonmagnetic systems, as was first done in
Ref.~\cite{orhanFirstprinciplesHubbardHund2020}, to 
calculate $U$ and $J$ simultaneously
(without any added approximation) using a single set
of finite-difference perturbations. This was recently
demonstrated to be an exact approach 
(always only for nonmagnetic systems) also in the context of
the conventional `self-consistent field' linear-response 
method for $U$, where it was termed the `gamma' 
method~\cite{lambertUseMathrmDFT2023}.
Here, perturbations of strength $\gamma$ are applied
uniformly to the spin-up channel of the subspace only,
in practice by setting $\alpha = \beta = \gamma / 2$
in the conventional notation for the linear response
Hubbard U. One can show that $\gamma$ is, in this regime,
a  parameter for both the numerators and denominators
of Eq.~\ref{eqn:basicUandJ}, so that a single regression
each, for $U$ and $J$, is sufficient, based on a 
common data set.

Bespoke NCPs were very painstakingly generated using the Rappe-Rabe-Kaxiras-Joannopoulos algorithm~\cite{rappeOptimizedPseudopotentials1990} as implemented in {the open-source pseudopotential generation software OPIUM (version 3.8)}
with a cutoff wavevector of 7.9 Ry$^{1/2}$ and 10 Bessel functions for each pseudo-orbital. In the construction of each pseudopotential, a neutral atomic reference configuration was solved using the j-averaged scalar relativistic scheme~\cite{grinbergTransferableRelativisticDiracSlater2000}
with the Perdew-Burke-Ernzerhof (PBE) functional~\cite{perdewGeneralizedGradientApproximation1996}. Non-linear core valence interactions were accounted for using a Louie-Froyen-Cohen-type partial core correction~\cite{louieNonlinearIonicPseudopotentials1982}. 

All DFT calculations were executed using the ONETEP (Order-N Electronic Total Energy Package) DFT code \cite{prenticeONETEPLinearscalingDensity2020,skylarisIntroducingONETEPLinearscaling2005,skylarisNonorthogonalGeneralizedWannier2002,oreganLinearscalingDFTFull2012}. ONETEP is, in principle, a linear scaling DFT code, which is achieved through truncation of the exponentially decaying tail of the density matrix
(this was not done here). The ONETEP code constructs the KS density matrix $\rho({\bf r},{\bf r}')$ from a set of localized basis orbitals, namely non-orthogonal generalized Wannier functions (NGWFs) $\{ \phi_{\alpha} \}$,
\begin{equation}
\rho({\bf r},{\bf r}')=\sum_{\alpha, \beta}\phi_{\alpha}({\bf r})K^{\alpha \beta}\phi_{\beta}({\bf r}'),
\end{equation}
where $K^{\alpha \beta}$ is the density kernel. Near-complete basis-set accuracy is achieved using a relatively small set of localized basis orbitals by minimizing the total energy of the system with respect to both $K^{\alpha \beta}$ and $\{ \phi_{\alpha} \}$.

All calculations were completed using the PBE \cite{perdewGeneralizedGradientApproximation1996} exchange-correlation (XC) functional at a cutoff energy of no less than $850$ eV. The convergence threshold of the root-mean-square gradient of the NGWFs and the electronic energy tolerance were set at $2\times 10^{-6}$ Ha $a_0^{3/2}$ and $1\times 10^{-4}$ eV/atom, respectively. Four NGWFs were assigned per sulphur atom and nine NGWFs were assigned per non-sulphur atom. For each atom, the NGWF cutoff radii were set to encapsulate 99.8\% of the norm of the respective pseudo-atomic KS wavefunctions computed by the pseudoatomic solver~\cite{ruiz-serranoPulayForcesLocalized2012} at the onset of the ONETEP calculation. A $1$,$728$ atom simulation cell comprising $6\times 6\times 3$ copies of the conventional cell was used in all calculations, at $\Gamma$ only. 
A Gaussian smearing of 0.1 eV was used for density-of-states
plotting.
Atomic positions and lattice constants were taken from the neutron diffraction results of Mangelis \textit{et al.}~\cite{mangelisUnderstandingOriginDisorder2019}.

\begin{table}[b]
\caption{\label{tab:table1}%
Hubbard $U$ and Hund's $J$ parameters computed via the simple $2\times 2$ minimum tracking linear response approach for CZTS and CZGS using orbitals as generated by the ONETEP pseudoatomic solver using bespoke NCPs.}
\def\arraystretch{1.3}
\begin{tabular}{cccc}
\hline\hline
\textrm{Material} & \textrm{Subspace} &
\textrm{$U$ (eV)}&
\textrm{$J$ (eV)} \\ \cmidrule{1-4}
& \cellcolor{Cu}{\hspace{7mm} \tcw Cu $3d$ \hspace{7mm}} & \hspace{2mm} 9.68 \hspace{2mm} & \hspace{2mm} 0.86 \hspace{2mm} \\
& \cellcolor{Sn}{\tcw Sn $5s$} & 2.50 & 0.67 \\
\multirow{-3}{*}{{\bf \hspace{7mm} CZTS} \hspace{7mm}} & \cellcolor{S}{\tcw S $3p$} & 4.74 & 0.57 \\ \hline
& \cellcolor{Cu}{\tcw Cu $3d$} & 10.02 & 0.89 \\
& \cellcolor{Ge}{\tcw Ge $4s$} & 3.11  & 0.79 \\
\multirow{-3}{*}{\bf CZGS} & \cellcolor{S}{\tcw S $3p$} & 4.65  & 0.57\\
\hline\hline
\label{hubbard_param_table}
\end{tabular}
\end{table}

The DFT+$U$ approach is primarily designed to improve the description of localized electronic states which are poorly treated by conventional local and semi-local XC functionals. {A topical question, however, is} which subspaces should be selected for treatment at the DFT+$U$-level? As opposed to arbitrarily selecting atomic $d$-states for treatment at the DFT+$U$-level, in this study we selected all atomic states that dominate the valence and conduction band edges. The reasoning behind this is two-fold; (i) the linear response technique is specifically designed for application with partially occupied states near the Fermi level, and (ii) correction of the local and semi-local XC functional{s' tendency to underestimate the bandgap} likely warrants the treatment of all atomic states which dominate the valence and conduction band edges, as is corroborated by the work of Wexler \textit{et al.}~\cite{wexlerExchangecorrelationFunctionalChallenges2020}, where application of Hubbard $U$ corrections to Cu $3d$, Zn $3d$,
and Sn $4d$ states failed to sufficiently open the bandgap of CZTS. By contrast, Nor \textit{et al.}~\cite{norInfluenceHubbardCorrection2024} reproduced the experimental bandgap through empirical tuning of the Hubbard corrective parameters applied to the Cu $3d$, Zn $3d$, Sn $5p$, and S $3p$ states, {albeit} at the loss of the predictive power of the technique.  

\begin{figure}[t]
\includegraphics[trim={0.1cm 0cm 0.05cm 0cm},clip,width=1.0\linewidth]{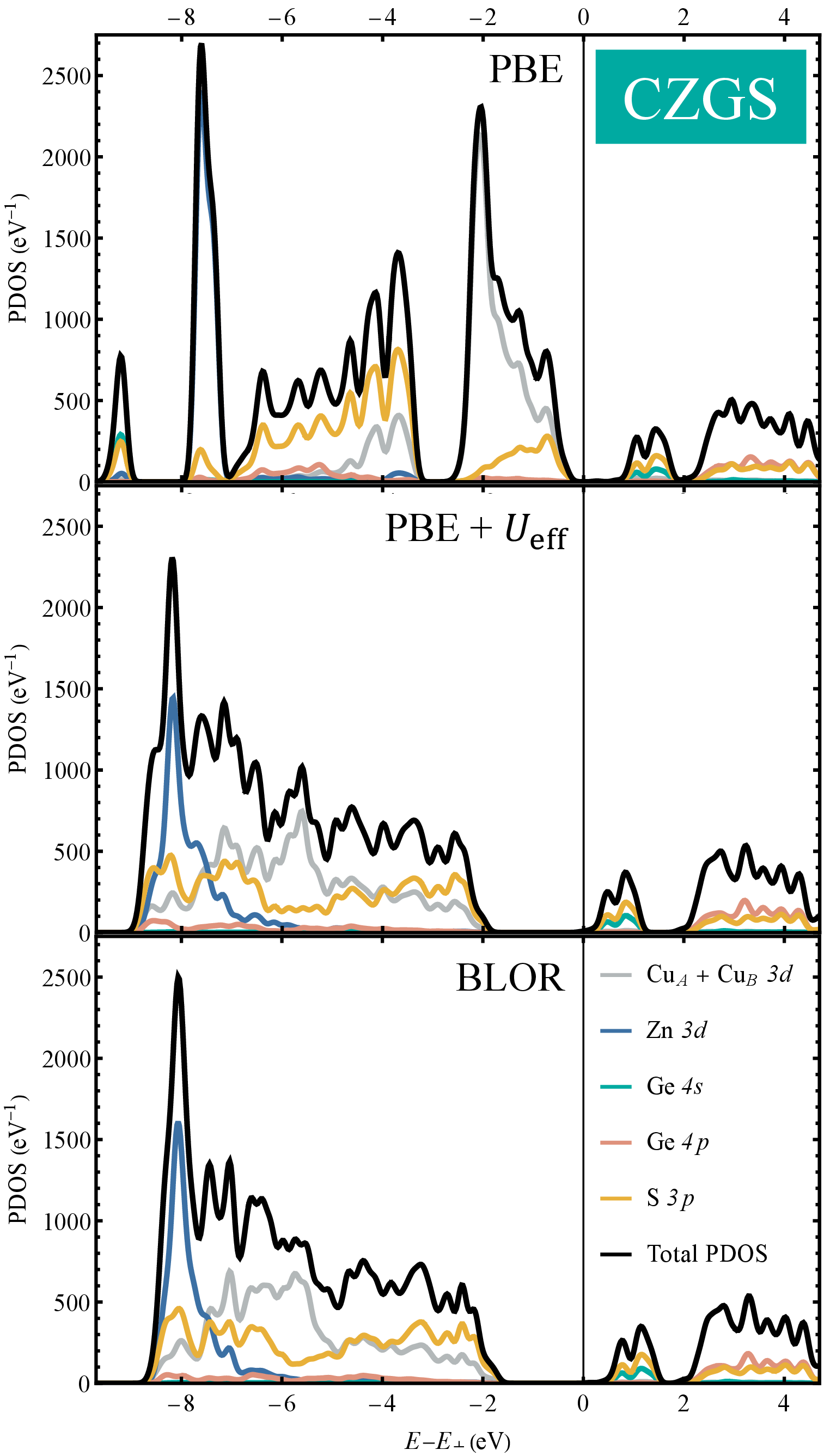}
\caption{Projected density of states (PDOS) of CZGS evaluated at the PBE (top), PBE+$U$$_{\rm eff}$ (middle), and PBE+BLOR (bottom) levels with a Gaussian broadening of 0.1 eV. The energy values are reported with respect to the mid-gap energy of the bare PBE calculation ($E_\perp$) to readily identify the effect of the Hubbard corrections on the valence and conduction bands.}
\label{fig:czgs_pdos}
\end{figure}

\begin{figure}
\includegraphics[trim={0.03cm 0.05cm 0.05cm 0cm},clip,width=1.0\linewidth]{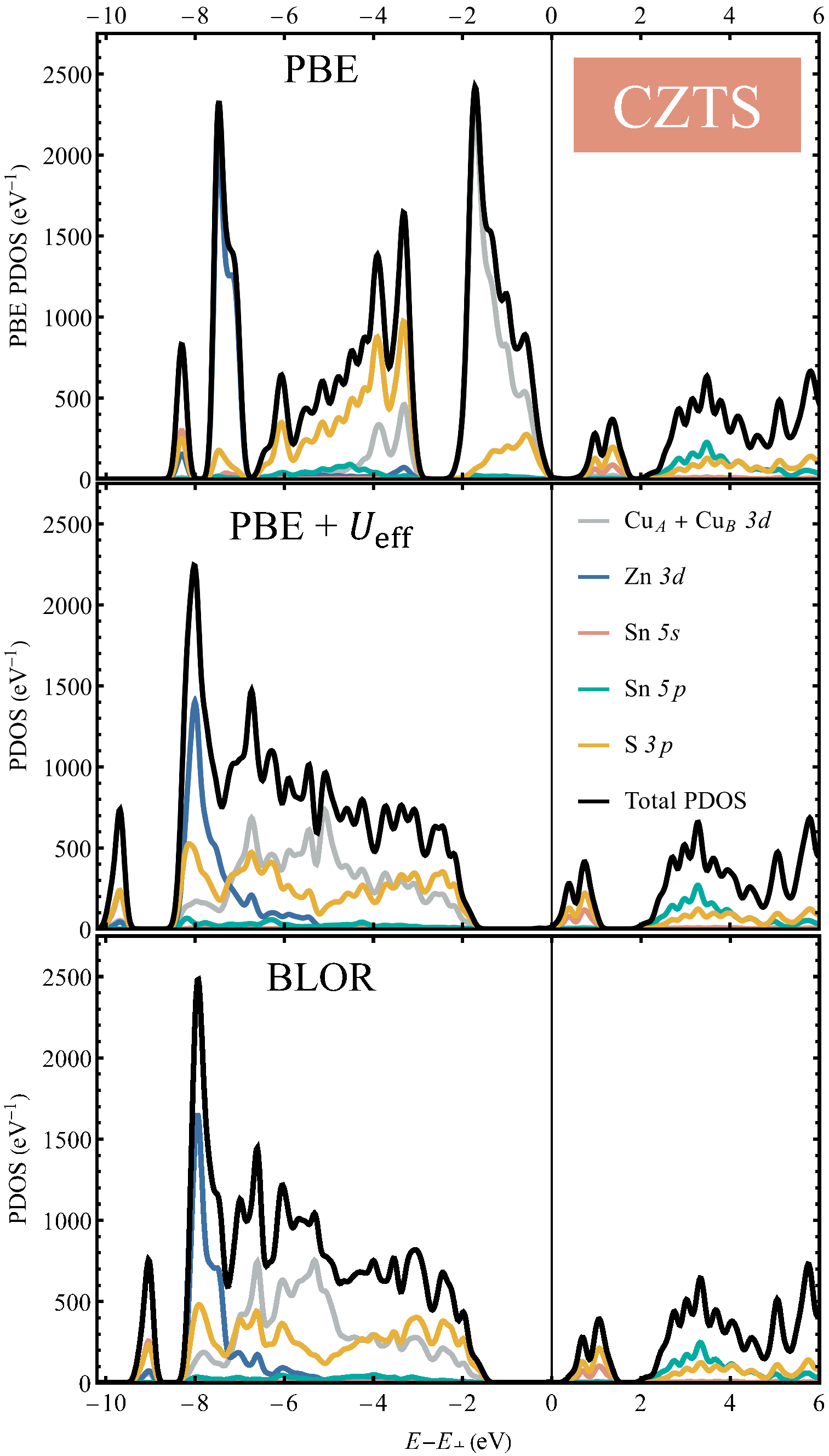}
\caption{PDOS of CZTS evaluated at the PBE (top), PBE+$U$$_{\rm eff}$ (middle), and PBE+BLOR (bottom) levels with a Gaussian broadening of 0.1 eV. The energy values are reported with respect to the mid-gap energy of the bare PBE calculation ($E_\perp$) to readily identify the effect of the Hubbard corrections on the valence and conduction bands.}
\label{fig:combined_czts_pdos}
\end{figure}

Fig.~\ref{fig:czgs_pdos} presents the projected density of states (PDOS) of CZGS evaluated at the bare PBE, PBE+$U_{\rm eff}$ and PBE+BLOR levels. An analogous plot for CZTS is presented in Fig~\ref{fig:combined_czts_pdos}. The valence and conduction band edges are dominated by Cu $3d$, S $3p$, and Ge $4s$ or Sn $5s$ (for CZGS or CZTS, respectively) states, and as such, these states  were selected for treatment at the DFT$+U$-level. The computed Hubbard $U$ and Hund's $J$ parameters are presented in Table~\ref{hubbard_param_table}. Small but non-negligible differences in the corrective parameters for the Cu $3d$ and S $3p$ atomic subspaces in CZTS compared to CZGS are observed. 

{Due to the near equivalent subspace occupancy matrices of the Cu atoms at the 2a and 2c Wyckoff positions, a single set of $U$ and $J$ parameters was used for both sites.  These $U$ and $J$ values were evaluated by applying perturbations to the $3d$ subspace of Cu at the 2a Wyckoff position. No Hubbard corrective parameters were evaluated for the Zn $3d$ subspace. By virtue of its absence from both the valence and conduction band edges the subspace would not be amenable to Hubbard parameter evaluation via linear response and for the same reason, reasonable Hubbard type corrections to the subspace would offer no significant correction to the bandgap.}

\begin{figure}[b]
\includegraphics[trim={0.2cm 0.3cm 0.2cm 0cm},clip,width=1.0\linewidth]{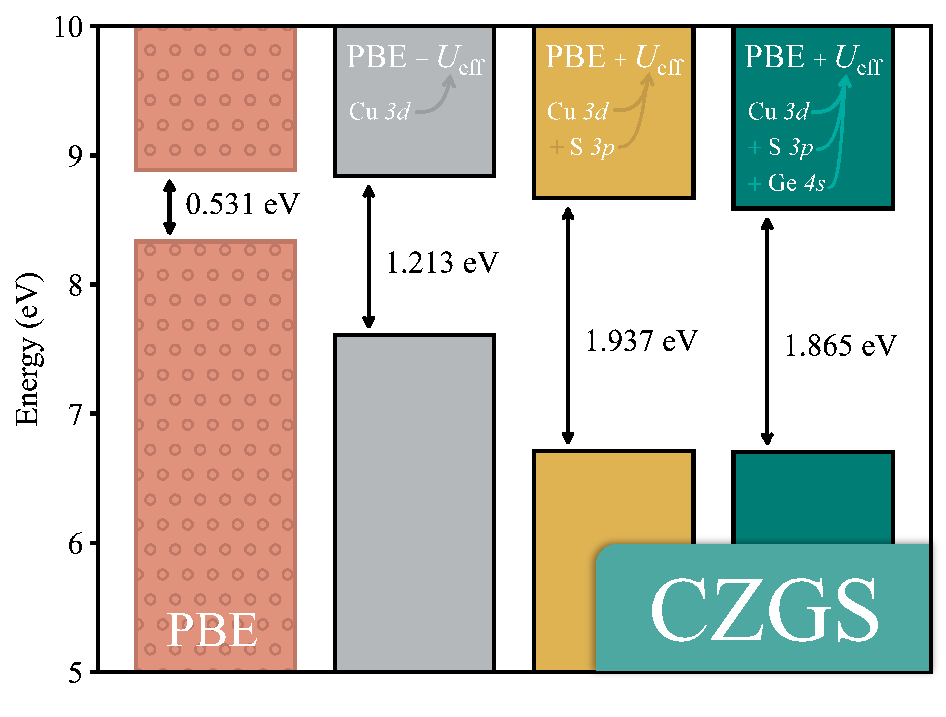}
\caption{The predicted bandgap of CZGS using the PBE XC functional with Hubbard corrections applied to an increasing number of atomic subspaces. The element labels refer to the atomic subspaces to which Hubbard corrections were applied, namely Cu-3d, S-3p and Ge-4s.}
\label{fig:bandgap_vs_subspace}
\end{figure}

The predicted fundamental bandgap of CZGS is presented in Fig.~\ref{fig:bandgap_vs_subspace} with an increasing number of atomic subspaces selected for treatment at the PBE+$U_{\rm eff}$ level. Here PBE+$U_{\rm eff}$ refers to the DFT+$U$ functional of Dudarev et al. with the effective Hubbard parameter $U_{\rm eff}$, set as the difference between the Hubbard $U$ and Hund's $J$ parameters presented in Table~\ref{hubbard_param_table}
(the calculated Hubbard $U$ is, most emphatically,
not already $U_{\rm eff}$~\cite{linscottRoleSpinCalculation2018}). The PBE approximation was applied as the base functional in all DFT+$U$ calculations. Treatment of the Cu-3d states alone did not sufficiently open the bandgap but this issue was rectified through the additional treatment of the S-3p states. The inclusion of Hubbard corrections on the Ge-4s states yields only a marginal change in the bandgap value, 
and the  addition of a positive $U$ on conduction band 4s states 
causing a small reduction in overall bandgap is an effect previously
observed in other materials.
Notwithstanding,  both the Cu(3d)+S(3p) and Cu(3d)+S(3p)+Ge(4s) options offer bandgaps in close agreement with the reported experimental value of 1.88~eV~\cite{khadkaStudyStructuralOptical2013}. 
These results corroborate our working hypothesis that reliable bandgap predictions, free from prior judgement regarding
subspace selection (there remains arbitrariness with
respect to projector shape), can be achieved through the application of Hubbard corrections to all atomic states that dominate the valence and conduction band edges. 

\begin{figure}[t!]
\includegraphics[trim={0.03cm 0cm 0.05cm 0cm},clip,width=1.0\linewidth]{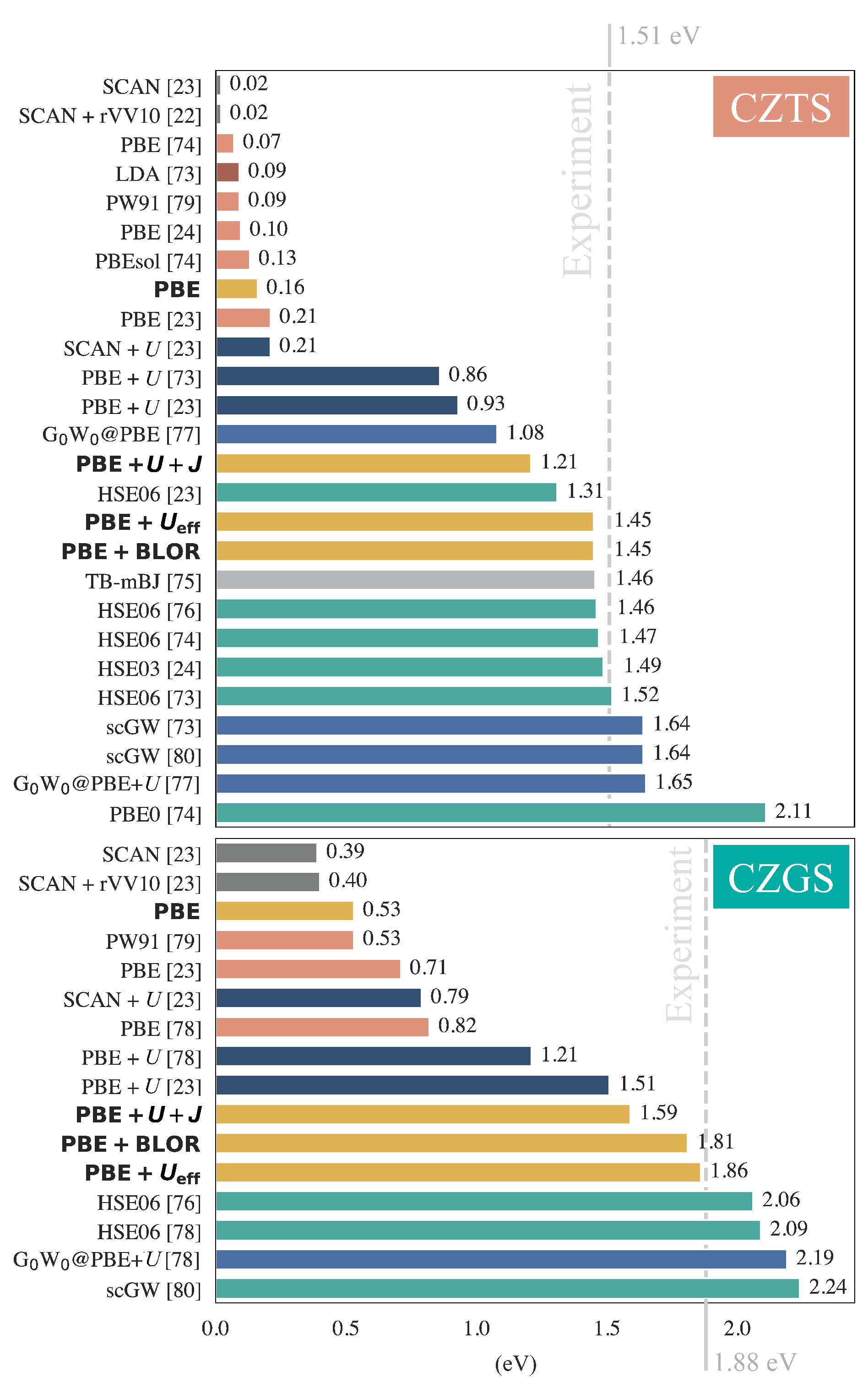}
\caption{Bandgaps of (top) CZTS and (bottom) CZGS evaluated at the PBE, PBE+$U_{\rm eff}$ and PBE+$U$+$J$ level with Hubbard corrections applied to the Cu $3d$, S $3p$, Sn $5s$, and Ge $4s$  atomic subspaces. A variety of bandgap predictions from the literature are also presented ~\cite{khadka2014structural,bottiBandStructuresCu2ZnSnS42011,khadkaStudyStructuralOptical2013,m.quennetFirstPrinciplesCalculations,n.dilshodDFTStudyStructure2022,paier$textCu_2textZnSnS_4$PotentialPhotovoltaic2009,parkStabilityElectronicProperties2018,wexlerExchangecorrelationFunctionalChallenges2020,zhangComparativeStudyStructural2011,zhangStructuralPropertiesQuasiparticle2012,chenElectronicStructureStability2009,korbel2015optical}, these were evaluated using a Local Density Approximation (LDA)~\cite{ceperleyGroundStateElectron1980} (dark pink bar)); semi-local approximations (PBE~\cite{perdewGeneralizedGradientApproximation1996}, PBEsol~\cite{perdewRestoringDensityGradientExpansion2008}, and PW91~\cite{Burke1998}—light pink bars); a meta-generalized gradient approximation, both with and without dispersion corrections (SCAN~\cite{sunStronglyConstrainedAppropriately2015} and SCAN+rVV10~\cite{sabatiniNonlocalVanWaals2013}—dark gray bars); hybrid functionals (PBE0~\cite{perdewRationaleMixingExact1996}, HSE03~\cite{heydHybridFunctionalsBased2003}, and HSE06~\cite{krukauInfluenceExchangeScreening2006}—green bars); the DFT+$U$ method~\cite{dudarevElectronenergylossSpectraStructural1998} (PBE+$U$ and SCAN+$U$—dark blue bars); and both the perturbative and fully self-consistent GW approximation (light blue bars)~\cite{hybertsenElectronCorrelationSemiconductors1986,brunevalEffectSelfconsistencyQuasiparticles2006}. Experimental values from UV-Vis absorption spectroscopy are also provided~\cite{khadkaStudyStructuralOptical2013,khadka2014structural}. Bandgaps calculated as part of the current work are bolded and have yellow bars.}
\label{fig:bandgaps}
\end{figure}

{Fig.~\ref{fig:bandgaps} presents the fundamental bandgap predictions for CZTS and CZGS evaluated at the PBE, PBE+$U_{\rm eff}$, PBE+$U$+$J$ and PBE+BLOR level with Hubbard corrections applied to all three atomic subspaces. The results are benchmarked against reference experimental optical bandgaps of Khadka et al.~\cite{khadkaStudyStructuralOptical2013,khadka2014structural} and previously computed bandgaps from the literature. Caution should be taken in comparing the reference experimental optical bandgaps to computed fundamental bandgaps as the two differ by the exciton binding energy. In order to ascertain a rough estimate of the exciton binding energy, both the fundamental bandgap and the first triplet excitation energies were computed at the PBE+$U_{\rm eff}$ level. The first triplet excitation energies of CZTS and CZGS were evaluated as 1.46 eV and 1.87 eV, which differ from their respective fundamental bandgaps by only 0.01 eV. This suggests a negligible exciton binding energy in both kesterite compounds, a result which is corroborated by the work of K\"{o}rbel et al.~\cite{korbel2015optical}, who reported negligible exciton binding energies having evaluated both the fundamental bandgap using the self-consistent GW approach and the optical bandgap via the Bethe-Salpeter equation. The absence of significant exciton binding effects allows a direct comparison to be made between the computed fundamental bandgaps and reference experimental optical bandgaps, cautioning of course that subtle differences can still arise due to the presence of defects~\cite{huang2013band} and finite grain-size~\cite{prabeesh2019czts}, zero-point renormalization~\cite{engel2022zero}, spin-orbit coupling~\cite{sabino2024impact}, and ambiguities in the extrapolation method used to ascertain the bandgap from UV-Vis absorption data~\cite{malerba2014czts}.}

Unsurprisingly, the bare PBE calculations vastly underestimated the bandgap for both kesterite structures. This failure, known more broadly as the bandgap problem~\cite{perdewDensityFunctionalTheory1985,borlidoLargeScaleBenchmarkExchange2019}, is already well documented in the literature and is related to the absence of a derivative discontinuity in conventional (semi-)local XC potentials with respect to electron count. Our PBE bandgap values are at least in close agreement with previous PBE results~\cite{paier$textCu_2textZnSnS_4$PotentialPhotovoltaic2009,wexlerExchangecorrelationFunctionalChallenges2020,m.quennetFirstPrinciplesCalculations,zhangStructuralPropertiesQuasiparticle2012}. Intriguingly, the poor bandgap predictions at the (semi-)local level~\cite{paier$textCu_2textZnSnS_4$PotentialPhotovoltaic2009,wexlerExchangecorrelationFunctionalChallenges2020,m.quennetFirstPrinciplesCalculations,zhangStructuralPropertiesQuasiparticle2012,bottiBandStructuresCu2ZnSnS42011,chenElectronicStructureStability2009} are not even partially alleviated through use of the SCAN meta-generalized gradient approximation~\cite{wexlerExchangecorrelationFunctionalChallenges2020}. Non-negligible bandgaps for CZTS can be achieved through use of the global hybrid PBE0~\cite{m.quennetFirstPrinciplesCalculations} or DFT$+U$-type corrections applied to the localized atomic $d$-states~\cite{wexlerExchangecorrelationFunctionalChallenges2020,bottiBandStructuresCu2ZnSnS42011,zhangStructuralPropertiesQuasiparticle2012}, however the results remain far from quantitative agreement with experiment. Furthermore, the use of the perturbative G$_0$W$_0$ method results in bandgaps that are exceedingly dependent on the chosen starting point~\cite{zhangComparativeStudyStructural2011,zhangStructuralPropertiesQuasiparticle2012}, be it a KS wavefunction computed at the PBE or PBE$+U$ level. Previous studies thus suggest that reasonable bandgap predictions for CZTS and CZGS semiconductors can only be achieved through computationally demanding techniques such as range-separated hybrids~\cite{paier$textCu_2textZnSnS_4$PotentialPhotovoltaic2009,wexlerExchangecorrelationFunctionalChallenges2020,parkStabilityElectronicProperties2018,m.quennetFirstPrinciplesCalculations,bottiBandStructuresCu2ZnSnS42011,zhangStructuralPropertiesQuasiparticle2012} or self-consistent GW approaches~\cite{bottiBandStructuresCu2ZnSnS42011,korbel2015optical}.

\begin{table}[b]
\caption{\label{tab:table2}%
Subspace averaged Hubbard corrective potentials in eV. The reported Cu $3d$ subspace values are for the  Wyckoff 2a position.}
\def\arraystretch{1.3}
\begin{tabular}{cccc}
\hline\hline
\textrm{Material} & \textrm{Subspace} &
\textrm{DFT+$U_{\rm eff}$}&
\textrm{BLOR} \\ \cmidrule{1-4}
& \cellcolor{Cu}{\hspace{7mm} \tcw Cu $3d$ \hspace{7mm}} & \hspace{2mm} -4.35 \hspace{2mm} & \hspace{2mm} -4.27 \hspace{2mm} \\
& \cellcolor{Sn}{\tcw Sn $5s$} & -0.70 & -0.37 \\
\multirow{-3}{*}{{\bf \hspace{7mm} CZTS} \hspace{7mm}} & \cellcolor{S}{\tcw S $3p$} &-1.72 & -1.20 \\ \hline
& \cellcolor{Cu}{\tcw Cu $3d$} & -4.50 & -4.43 \\
& \cellcolor{Ge}{\tcw Ge $4s$} & -0.94  & -0.63 \\
\multirow{-3}{*}{\bf CZGS} & \cellcolor{S}{\tcw S $3p$} & -1.68  & -1.17\\
\hline\hline
\end{tabular}
\end{table}

{Despite this}, our PBE+$U_{\rm eff}$ approach yields bandgaps in remarkably good agreement with experiment, deviating by only 0.06 eV and 0.02 eV from the CZTS and CZGS reference values, respectively. In both cases, the PBE+$U_{\rm eff}$ bandgap prediction outperforms even the self-consistent GW approach~\cite{brunevalEffectSelfconsistencyQuasiparticles2006,bottiBandStructuresCu2ZnSnS42011,korbel2015optical}. The inclusion of inter-spin corrections via the extended PBE+$U$+$J$ functional degrades the predicted bandgaps by about 0.3~eV. To understand this discrepancy, consider a system with Hubbard corrections applied to one Hubbard manifold which is  at full subspace occupancy. To first-order in perturbation theory, Dudarev's functional will, in the limit of full subspace occupancy, shift a band which projects perfectly onto the Hubbard manifold by $-(U-J)/2$. The DFT+$U$+$J$ functional will shift the same band by $-(U-3J)/2$. If the valence band edge projects perfectly onto the Hubbard manifold, the bandgap opening offered by Dudarev's functional and the DFT+$U$+$J$ functional differs by $J$~eV. In contrast, the BLOR functional will, like Dudarev's functional, shift the band by $-(U-J)/2$. The Cu $3d$ subspace in CZTS and CZGS is near full occupancy, thus unsurprisingly Dudarev's and BLOR functional predict similar bandgaps for the two kesterite compounds while the DFT+$U$+$J$ functional offers a reduced bandgap in both cases. {However, it is worth emphasizing that the exact numerical agreement between the PBE+$U_{\rm eff}$ and PBE+BLOR bandgaps for CZTS is merely fortuitous as the two approaches offer quite distinct corrections to the potential at the sulfur and tin/germanium sites, as indicated by the subspace averaged Hubbard corrective potentials in Table~\ref{tab:table2}. While the two DFT+$U$ functionals offer similar corrections to the Cu $3d$ subspace, the BLOR functional offers a smaller correction to the S $3p$ and Sn $5s$/Ge $4s$ subspaces in both materials.} The underperformance of the DFT+$U$+$J$ functional in this study, is in line with recent findings of the extended corrective functional's failure to improve the DFT+$U$ adiabatic energy differences in spin-crossover transition metal complexes~\cite{macenultyBenchmarkingTotalEnergies2025},
but we note however that it has been shown to handle
a test-set of nonmagnetic transition metal oxides very
well~\cite{lambertUseMathrmDFT2023}. {Despite the promising success of the BLOR functional in predicting the bandgaps of CZTS and CZGS, further testing and possibly refinement of the corrective functional will be needed before the authors consider advocating for the method's widespread adoption on solids. }

\begin{table}[b]
\caption{\label{tab:table3}%
The valence bandwidth and energy gap between the first and second valence bands (${\rm VB} \rightarrow {\rm VB}-1$), evaluated via application of the DFT+$U_{\rm eff}$ and BLOR functionals.}
\def\arraystretch{1.3}
\begin{tabular}{cccc}
\hline\hline
   & Corrective \hspace{1mm} &
\textrm{Valence \hspace{1mm} }&
\textrm{${\rm VB} \rightarrow {\rm VB}-1$ } \hspace{1mm} \vspace{-1mm}  \\  \multirow{-2}{*}{{  Material} \hspace{1mm}}    & Functional \hspace{1mm} &
\textrm{Bandwidth (eV) \hspace{1mm} }&
\textrm{Gap (eV) } \hspace{1mm} \\ \hline
\multicolumn{1}{c}{\cellcolor{Sn}} & DFT+$U_{\rm eff}$ & 6.75 & 1.13 \\
 \multirow{-2}{*}{{\cellcolor{Sn}{\tcw\bf  CZTS}} \hspace{0mm}}  & BLOR &   6.71 & 0.76 \\ \hline
 \multicolumn{1}{c}{\cellcolor{Ge}} & DFT+$U_{\rm eff}$ & 7.06 & 1.64 \\
 \multirow{-2}{*}{{\cellcolor{Ge}{\tcw\bf  CZGS}} \hspace{0mm}} &BLOR & 7.02  & 1.30  \\ 
\hline\hline
\end{tabular}
\end{table}

The CZTS PDOS evaluated at the bare PBE, PBE+$U_{\rm eff}$ and PBE+BLOR levels are presented in Fig.~\ref{fig:combined_czts_pdos}. Despite the significant differences in the Hubbard corrections to the potential at the sulfur and tin sites, Dudarev's functional and the BLOR functional offer remarkably similar corrections to the PDOS of CZTS. The energies are reported with respect to the mid-gap energy of the bare PBE calculation for ease of comparison. Application of Hubbard-type corrections to the Cu $3d$, S $3p$, and Sn $5s$ states results in a bandgap opening primarily due to a significant lowering in the energies of the occupied bands of Cu $3d$ and S $3p$ character. Intriguingly, the lowering in energy of the localized Cu $3d$ states results in increased hybridization with the broader S $3p$ states. Therefore, rather counterintuitively, the application of Hubbard-type corrections results in more energetically delocalized Cu $3d$ bands in CZTS. The conduction band edge is predominantly of S $3p$ and Sn $5s$ character, both with and without Hubbard-type corrections. The application of the Hubbard-type corrections results in a marginal lowering in energy of the unoccupied $3p$ and Sn $5s$ states. The Sn $5p$ in addition to the zinc and sulphur $s$ states contribute significantly to the conduction band beyond 2 eV, however, of these, only the Sn $5p$ contribution was individually plotted to avoid overcrowding the graph. 
Moreover, we note that as the NGWFs in ONETEP are optimized
to minimize the total energy, so they are not optimized
as standard to form a complete basis for the unoccupied states.
Thus, the conduction band PDOS shown should be interpreted
as qualitatively but not quantitatively accurate for the 
approximate functional at hand. {By contrast, for a given approximate functional a quantitatively accurate valence band can be achieved with the ONETEP code. The valence bandwidth and energy gap between the first and second valence bands are reported in Table~\ref{hubbard_param_table}. In both materials, the DFT+$U_{\rm eff}$ and BLOR functionals predict remarkably similar values for the valence bandwidth, but the BLOR functional offers a reduced valence `second' energy gap compared to the DFT+$U_{\rm eff}$ functional. }

\begin{figure}[b]
\includegraphics[trim={8.9cm 3.0cm 0.0cm 9.5cm},clip,width=1.25\linewidth]{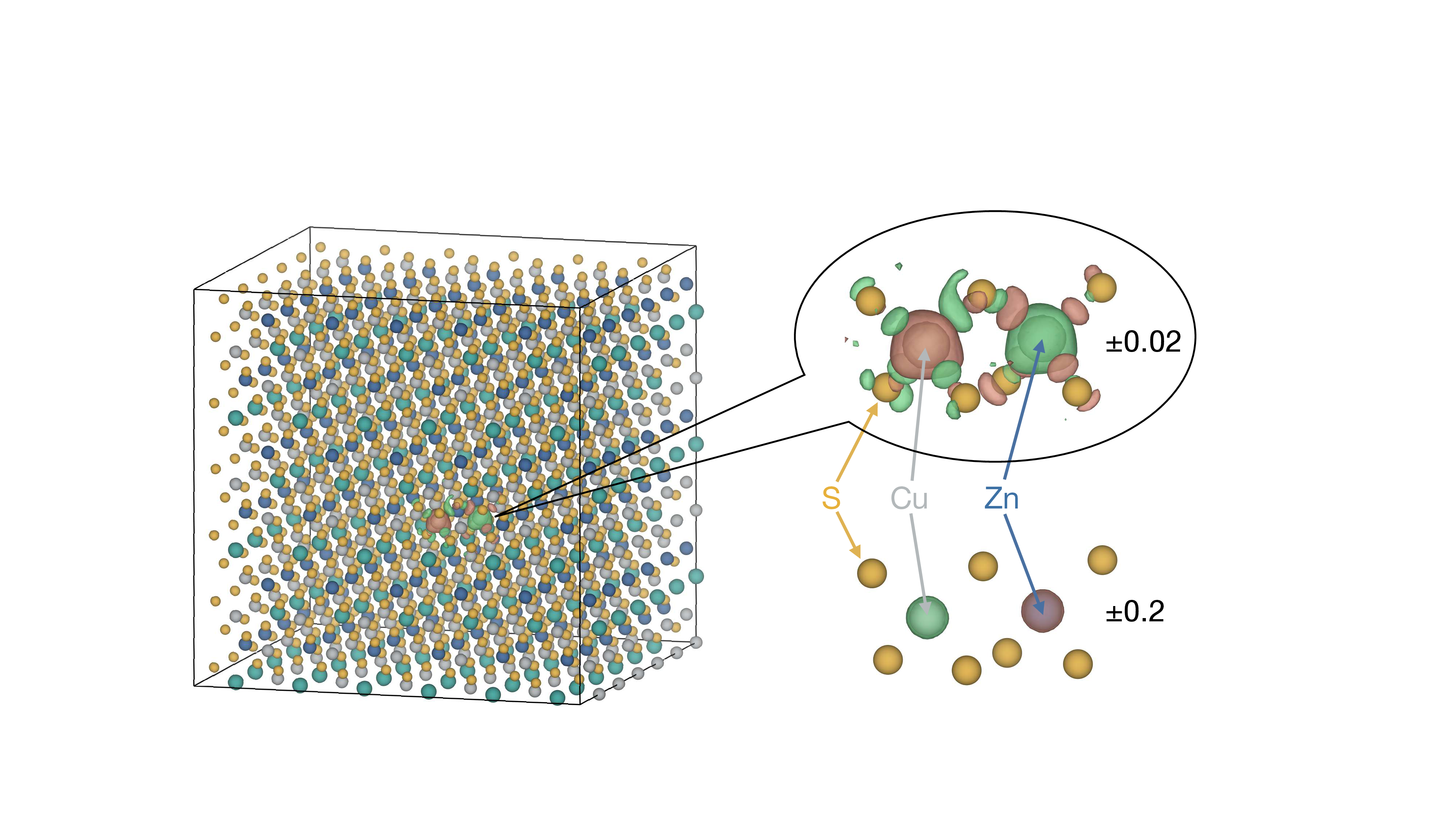} 
\caption{\label{DefectDensity} The Cu-Zn antisite pair defect density at an isosurface value of $\pm 0.02$ \AA$^{-3}$ and $\pm 0.2 $ \AA$^{-3}$~\cite{mommaVESTAThreedimensionalVisualization2008}, evaluated as the density difference between the defect-harboring and pristine structures at the PBE+$U_{\rm eff}$ level. Cu is shown in gray, Zn in blue, Sn in dark green, and S in yellow. Density increases are shown in brown and density decreases in light green.}
\end{figure}

The linear scaling ONETEP code readily enables DFT calculations on large simulation cells; this, combined with the very satisfactory PBE+$U$$_{\rm eff}$ bandgap predictions, offers a computational approach which is ideally suited for the prediction of crystallographic defects. Previous computational studies suggest a wide variety of possible defects may form in CZTS under experimental conditions~\cite{chenIntrinsicPointDefects2010,kimIdentificationKillerDefects2018,wexlerExchangecorrelationFunctionalChallenges2020, xiao2015intrinsic,maedaFirstPrinciplesCalculations2011,kumarStrategicReviewSecondary2015,wangDefectsKesteriteMaterials2024,schorrPointDefectsCompositional2019}.
Thanks to their ability to act as free carrier traps and non-radiative recombination centers~\cite{chenIntrinsicPointDefects2010}, the presence of point defects can have a significant impact on solar cell efficiencies. Identifying and characterizing point defects in kesterite-based semiconductors is thus of central importance to improving device performance. The results of neutron diffraction experiments suggest a partial disorder in the cation sub-lattice, with Cu$_{\rm Zn}$ and Zn$_{\rm Cu}$ anti-sites forming at the Wyckoff 2d and 2c positions, respectively~\cite{mangelisUnderstandingOriginDisorder2019,schorrNeutronDiffractionStudy2007,schorrCrystalStructureKesterite2011}. In this study, the formation of the Cu-Zn anti-site pair defect (Cu$_{\rm Zn}$+Zn$_{\rm Cu}$) was investigated at the PBE+$U_{\rm eff}$ level. The Cu-Zn anti-site pair defect is presented in Fig.~\ref{DefectDensity} based on the density difference of the defect-harboring and pristine structures at the PBE+$U_{\rm eff}$ level. There is a sharp decrease in the electron density in the immediate vicinity of the Cu site, an increase in the electron density at larger radii and a subsequent decrease in the electron density in the Cu-S bonding region. The reverse is true at the Zn site. This effect is largely due to the additional 4s electron in Zn compared to Cu with the subtle changes in charge density at larger radii and in the transition metal-sulfur bonding region being simply due to charge compensation effects. The defect density plot suggests that both the Cu$_{\rm Zn}$ site and Zn$_{\rm Cu}$ site remain charge neutral and together form a charge neutral defect complex.  This is corroborated by the Mulliken atomic (NGWF based) population analysis, where both the standard Cu site and the defective Cu$_{\rm Zn}$ site have a total occupancy of 11.1, while the occupancy at the standard Zn site and the defective Zn$_{\rm Cu}$ site are both 11.7.

\begin{figure}
\includegraphics[trim={0.2cm 0.3cm 0.2cm 0cm},clip,width=1.0\linewidth]{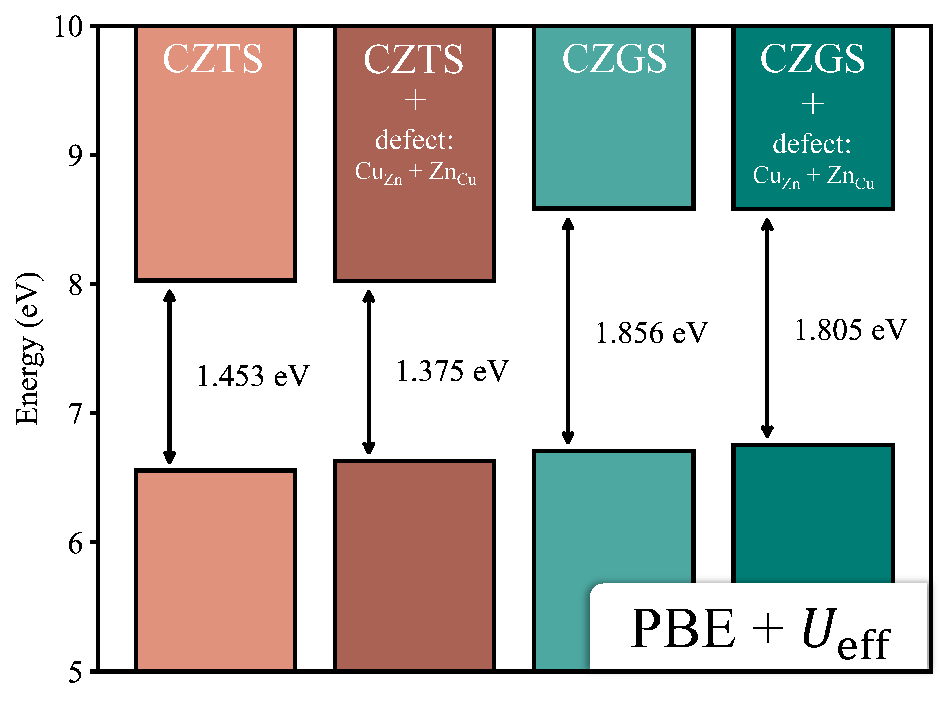}
\caption{{The bandgaps (interstitial text) of CZTS (pink) and CZGS (teal), both with (darker bars) and without (lighter bars) the charge-neutral anti-site pair defect (Cu$_{\rm Zn}+$Zn$_{\rm Cu}$), evaluated at the PBE+$U$$_{\rm eff}$ level. Valence band represented by lower tier of bars, and conduction band represented by the upper bars.}}
\label{fig:defect_bandgaps}
\end{figure}

In Fig.~\ref{fig:defect_bandgaps}, the bandgaps of pristine CZTS and CZGS are compared to the equivalent structures harboring a Cu-Zn anti-site pair defect. The defect geometry was prepared by simply swapping the chemical character of neighboring Cu and Zn sites without running a geometry relaxation. As shown in Fig.~\ref{fig:defect_bandgaps}, no defect levels are observed within the gap; rather the inclusion of the anti-site pair defect results in a slight decrease of the band-gap and a corresponding increase in the energy of the valence band maximum.  The defect-formation energy of the charge-neutral, stoichiometry-preserving anti-site pair defect is independent of the chemical potential and can thus be readily evaluated as the energy difference between the pristine and defect-harboring crystallographic structures. The frozen-ion defect-formation energies in CZTS and CZGS were evaluated as 0.580 eV and 0.505 eV, respectively. 

In conclusion, \textit{in situ} Hubbard parameters determined via minimum-tracking linear response were deployed through a series of common DFT$+U$ functionals for the prediction of the electronic structure of kesterite-based photovoltaic materials. The conventional DFT$+U$-type functional of Dudarev \textit{et al.} predicted bandgaps in remarkably close agreement to experimental reference values, even marginally outperforming the self-consistent GW approach. As opposed to improving the DFT$+U_{\rm eff}$ approach, the extended DFT$+U$+$J$ functional introduces a systematic underestimation of the bandgap of CZTS and CZGS, this effect can be ameliorated through the use of the alternative Hubbard $U$ plus Hund's $J$ extended functional of Burgess Linscott and O'Regan. In contrast to conventional understanding of the DFT$+U$ approach, the inclusion of Hubbard $U$ corrections on $3d$ transition metal sites results in a more energetically delocalized electronic structure for this particular material class. The DFT$+U_{\rm eff}$ approach was also deployed for the prediction of defects in CZTS and CZGS. A low frozen-ion defect-formation energy of 0.580 eV and 0.505 eV was predicted for the Cu-Zn anti-site pair in CZTS and CZGS, respectively.

This research was supported by Taighde-Éireann under Grant No. GOIPG/2020/1454 and under Prime Award No. 12/RC/2278$\_$P2, the latter of  which is co-funded by the European Regional Development Fund. {LM and DDO'R further acknowledge  the support of Trinity College
Dublin though its Provost PhD Awards.} The authors wish to acknowledge the Irish Centre for High-End Computing (ICHEC) for the provision of computational facilities and support. Calculations were also performed on the Boyle cluster maintained by the Trinity Centre for High Performance Computing. This cluster was funded through grants from the European Research Council and  Taighde-Éireann -- Research Ireland.

\bibliography{main}

\end{document}